# Smart Communities Internet of Things

Klara Nahrstedt, Daniel Lopresti, Ben Zorn, Ann W. Drobnis, Beth Mynatt, Shwetak Patel, and Helen V. Wright

January 12, 2016

Version 1

## Motivation

Today's cities face many challenges due to population growth, aging population, pedestrian and vehicular traffic congestion, water usage increase, increased electricity demands, crumbling physical infrastructure of buildings, roads, water sewage, power grid, and declining health care services [13], [14]. Moreover, major trends indicate the global urbanization of society, and the associated pressures it brings, will continue to accelerate [23]. One of the approaches to assist in solving some of the challenges is to deploy extensive IT technology. It has been recognized that cyber-technology plays a key role in improving quality of people's lives, strengthening business and helping government agencies serve citizens better.

## White Paper Goals

In this white paper, we discuss the benefits and challenges of cyber-technologies within "Smart Cities", especially the IoT (Internet of Things) for **smart communities**, which means considering the benefits and challenges of IoT cyber-technologies on joint *smart cities physical infrastructures* and their *human stakeholders*. To point out the IoT challenges, we will first present the framework within which IoT lives, and then proceed with the challenges, conclusions and recommendations.

## Relations to Existing Efforts

Many "Smart Cities" projects represent the aggregations of such cyber-technologies to assist in solving cities' challenges and within these cyber-technologies, a major impact area is IoT (Internet of Things) [1, 2]. There are many pilot "Smart Cities" projects underway world-wide, including Barcelona, Chicago [8], Singapore [1], Boston, Beijing [3], Nairobi, and others [6]. For example, the "Smart Nation" project in Singapore aims to enable safer, cleaner and greener urban living, more transport options, better care for the elderly at home, more responsive public services and more opportunities for citizen engagement [1].

There is related work discussed, for example, at the ANSI Smart and Sustainable City Events [15], the IEC SEG 1 – Systems Evaluation Group 1 on smart cities, the ISO/IEC JTC 1/SG 1 – Study group on smart cities, and the ITU-T SG 5 – Focus group on smart and sustainable cities [2]. Related *workshops* like the ANSI 2013 workshop discuss how to leverage innovations in urban informatics to drive improvements of smart grid, green building, energy and water use, waste management and transportation. Also, *standard committees* such as the various IEC, ISO, and ITU groups discuss computing and communication interfaces, platforms and service for interoperability among smart city technologies.

Many of the industry-oriented IoT white papers concentrate on *smart city applications*, for example, smart grid meters, smart transportation [16], smart lights [11], *economical benefits and risks analysis* of IoT, and *market shares* of IoT [10, 12] because industrial vendors want to be able to sell a robust solution for the city (e.g., [10, 11, 12, 16]). Major events have arisen to bring together providers and purchasers of IoT technologies (e.g., [24]). Our white paper reflects the views of members of the computer science research community, as distinct from, and complementary to, other groups who are contributing to the development and deployment of smart city technologies.

### Smart Communities Framework

Smart communities are a collection of interdependent human-cyber-physical systems, where IoT represents the sensing and actuating cyber-infrastructure to estimate the state of human and physical systems and assist in adapting/changing these systems. Smart cities are often classified along multiple dimensions:

(1) **IoT technological (cyber) workflow dimension** including (a) sensory "things" development and deployment, (b) connection of "things", (c) digital data collection from "things", (d) processing, aggregation, analytics of correlation of data according to human-physical models and urban domains/applications, (e) comprehension of data and findings, (f) creation of new services and actions, and (g) actuation of "things" to gain new data. The technological workflow represents a full loop of digital data life cycle from data capturing, monitoring, to collection, processing, analysis and feedback to the cyber-system according to physical or human models.

(2) **IoT urban domain and application dimension** including (a) urban mobility (transportation system), (b) health care, (c) utilities (e.g., smart grid, water, gas), (d) urban living (smart home technologies), (e) public services (incident reporting), (f) safety and security (police, first responders), (g) sustainability.

(3) **IoT stakeholders dimension** including *elected/appointed officials* as decision makers to select usage of urban systems and to ensure economic development and cost-effective usage of municipal resources; *city workers* as the implementers of decisions made by officials; *citizens* as beneficiaries of smart city services; and *vendors / developers / entrepreneurs* as providers of smart city hardware, software, and services, partners in economic development. These stakeholder roles are unique to the smart communities IoT problem because, for example, elected officials **interpret and act upon IoT data**, hence IoT data collection, analytics, and comprehension research must provide high accuracy solutions, based on short training time, to enable fast and high fidelity decision making process; city workers are vital to the **success and failure of IoT deployment** across communities, hence IoT sensing, actuation, and connectivity research must provide scalable and easy-to-deploy solutions; citizens support **IoT infrastructure investments**, hence IoT development, deployment, and usage research must be connected to IoT economics; and vendors have vested interests in seeing their **IoT solutions chosen over competing products**, which requires research in creation of new IoT services and applications.

**Smart Communities IoT Challenges**

We will discuss the IoT challenges along the technological (cyber) workflow with clear examples from various urban applications and stakeholders. From the cyber workflow point of view, IoT systems must satisfy requirements such as real-time, robustness, reliability, resilience, privacy and security to have a long-term impact in our cities. These IoT system requirements (e.g., real-time, robustness, security) represent major challenges for the cyber-workflow stages (e.g., data collection, analytics, comprehension, actuation stages) if we want these requirements to be satisfied in an end-to-end manner and ensure also a seamless integration into existing communities' testbeds and environments.

*Sensor Development and Deployment:* The IoT challenges span multiple hardware and software challenges. (a) major need to build efficient and effective sensors that will best serve the various urban applications. For example, in health care, **bio-inspired camera sensors** are needed such as imaging sensors developed at Washington University St. Louis, inspired by the mantis shrimp, and used for early cancer detection and optical neural recordings without the use of a fluorescence marker [16], and in utility applications such as water sewage, to enable **resilient sensors** for flooding detection and prevent flooding of basements in Chicago [6], we need sensors that operate in difficult environmental conditions and that are affordable to install and maintain over macro periods of time. (b) major need to provide near-optimal and cost effective deployment of sensors. For example, to identify and ease urban mobility problems, **identification of major points of interests** is needed to place mobility and congestion-detection sensors. (c) major need to provide real-time and trusted data capturing. For example, in smart homes, utility companies need to ensure that reported data from smart meters are **trusted** to provide accurate pricing signals to customers [18,19].

*Connection of "Things"*: The IoT challenge regarding connectivity of things spans multiple networking challenges. (a) major need for **seamless diverse wireless technologies** to enable extraction of a collection of sensory data that might be placed in dense or sparse topologies. For example, in case of urban vehicular mobility, congestion-detection sensors on highways might be placed sparsely, where inside city center the sensors might be placed in dense formation. Hence, we may utilize cellular networks for sparse placement of sensors vs WiFi wireless networks for dense placement of sensors. (b) major need for **latency, bandwidth and interference management** to enable connection of things in crowded and interfering scenarios. (c) major need for **context-aware connectivity of things** to enable energy-efficient networking of devices. For example, if we use mobile smartphones as urban human mobility sensors, understanding when and where to connect phones in energy-efficient manner is of great importance.

*Data Collection from Things*: The IoT data collection challenge spans **multiple routing, privacy and security challenges**. (a) major need for sensor network routing protocols that work in **delay-sensitive** and **delay-tolerant** manner. For example, in smart grid utility application, smart meter data collection may be done by routing protocols over diverse wireless and wired networks. It is very important to understand the SLAs (Service level agreements) how data are being routed and relayed, by whom, when, where, and in what form. All this information is needed to understand the delays during the data collection process. (b) major need for **privacy and security during data collection** since if data are being relayed, various entities, intercepting traffic, could infer the users' behaviors (e.g., when a person is at home), electricity usage, etc.

*Analytics of Data:* The IoT data analytics challenge spans multiple machine learning, data management, data mining, data processing challenges. (a) major need for **machine learning algorithms over large data sets** to find meaningful domain-specific insights from the data. For example, in case of smart grid utility domain, it is important for utilities to gain a strong understanding of electricity usage for the city and neighborhoods to assess further needs, load balancing and load shedding. (b) major need for **scalable, distributed and parallel data management and storage systems** to enable fast and efficient data access, data retrieval and data processing. For example, in case of a water system utility, when flooding is detected from collected data, city officials, city workers, citizens and visitors need to react fast either to inform public about upcoming events, and/or organize evacuations and major actions to protect citizens and to react from the side of citizens.

*Comprehension of Findings:* The IoT comprehension challenge spans multiple decision and control algorithms challenges. (a) major need for **decision algorithms** to make the right decision when analysis shows certain behaviors and findings. The decision algorithms might be at the **cyber level, human level and at the physical level**. For example, in case of the water utility flooding scenario, if flooding potentials are detected, the cyber level decision might be to increase the frequency of sensing to get more detailed situation awareness, the human level decision might be to inform the public and ask city workers to put up barriers, and at the physical level the decision might be to establish sand barriers. (b) major need for control algorithms to run different **"what if" scenarios** against different cyber-human-physical models and understand the interdependencies among different infrastructures. For example, in case of the flooding scenarios, interconnection with other cyber-physical infrastructures will be important since flooding might impact transportation infrastructure, healthcare infrastructure, and other public services.

*Creation of New Services and Actions:* The IoT challenge spans multiple **cyber-infrastructure challenges** including (a) **new configurations** of computing and networking infrastructures, (b) **inclusion of new sensors** (if some previous sensors were insufficient to detect certain situations), (c) **deployment of new policies and services** that will execute and react to decisions and control resulting from comprehension of data analytics, and (d) the **need to incrementally deploy and test new infrastructure**.

*Actuation of Things:* The IoT actuation challenge spans feedback challenges which may impact human level feedback and actuation, cyber-level feedback and actuation, and physical-level feedback and actuation. (a) major need to have **sensors/actuators** architecture in place which can take **feedback and change behaviors** if it is sensing frequency, type of data or functional behaviors. (b) major need to have **adaptive algorithms and policies** that can react to changing conditions if there are cyber-components, mechanical/physical components or human components.

**Conclusions**

In summary, Smart Communities IoT capabilities have a very promising and important place in solving the growing communities' challenges. However, to enable these IoT capabilities fully, trustworthy algorithmic and system designs must be integral parts of the overall Smart Community framework (and all of its dimensions) which includes not only the IoT cyber-systems challenges as we discussed above, but also the physical and human systems, understanding their models, constraints and characteristics.

With advanced IoT capabilities, there are tremendous opportunities to improve and impact the quality of life of urban communities [2] if we can solve the IoT technology evolution going from "**measured**" pervasive sensor networks throughout a city, "**networked**" node connections through low-cost communications, "**managed**" real-time analysis and control of city systems, to "**integrated**" systems which have been traditionally isolated within and across cities, and "**smart**" Software-as-a-Service citizen services, applications and management tools.

We conclude that to realize these opportunities, several important factors will need to be in place:

First, **funding of research, development and deployment (RD&D) of advanced IoT technologies and data tools** as discussed above will be the most important factor of the Smart Communities realization success. This is a problem since many cities do not have the needed funding for IoT RD&D. This means that industry players will want to step in to make the capital investment which will not be cheap. Major industry involvement in IoT RD&D for smart communities will present a tension between services provided to citizens and profit to be made by the providers. Hence, local, regional, state and federal governments need to provide the initial RD&D investments of IoT technologies and (a) work closely with *city-related organizations* to understand the value of IoT technologies to the city, and the resources needed for implementation and deployment via testbeds and experimental IoT-zones, and (b) work closely with groups of private industry partners to establish private-public partnerships and explore an incremental but massive IoT deployment that would not fully disrupt the community. Cities are already built, hence we cannot build a city from scratch with IoT-based cyber-infrastructures as first-class objects. IoT will need to be embedded in an incremental fashion.

Second, different "Smart Communities" sectors and services will need to **integrate their IoT cyber-infrastructures**. The current status is unsatisfactory. Many cities have their services and corresponding cyber-infrastructures (digital computing and communication) operated in silos and in an environment of competing interests. One example would be the usage of different wireless network technologies among different city service providers, e.g., the fire and police departments. The present tension is when city officials want to share "community" cyber-infrastructures to offer citizens'-oriented services and service providers want to operate in their own spaces for security, business or mission reasons. The integration across city service providers is needed to enable seamless service access anywhere and anytime.

Third, CS researchers, technologies and other stakeholders of Smart Communities will need to consider the **cost of deployment**. There are many IoT-enabled technologies, e.g., in transportation, which allow advanced transportation capabilities to lower the gas emissions, or to decrease number of accidents, or to optimize number of snow-plowing trucks and city will be excited to deploy them, but the city does not have funds. Therefore, with the government funding, smart communities need to consider how to create IoT zones (regions) and living laboratories to give stakeholders cyber-infrastructures and demonstrate ideas. For example, the SCOPE project (Smart-city Cloud-based Open Platform and Ecosystem), led by Boston University [7], provides access to Massachusetts Open Cloud and big data technologies to improve transportation, energy, public safety, asset management and social services in the City of Boston and across Massachusetts.

Forth, **security and privacy of IoT technologies** need to be become first-class objects and an integral part of the IoT RD&D, so that stakeholders have trust when using and relying on city-and-community-related services and tools. The IoT technologies and smart communities' cyber-infrastructures will not be broadly accepted without trustworthy solutions. But, the IoT trustworthy capabilities must go hand-in-hand with being real-time, have reliable performance, and being easy use, otherwise, the IoT solutions will not be accepted either.

Fifth, **data accessibility and data sharing** is a major factor. There is a present tension when city officials want to share "community" data to offer citizens'-oriented services and service providers want to own and monetize the data. The data sharing across different city service providers is needed to enable seamless data access, but with it must come privacy-preserving IoT technologies and cost-models that would yield commercial profit. Regarding this factor, the National Science Foundation is taking steps by establishing the NSF Big Data Regional Innovation Hubs, where Big Data for Smart Communities is an integral part of the each Hub [9].

**Recommendations**

There is a **major urgency of funding of Computer Science (CS) basic research, development and deployment** to develop novel IoT solutions and their related cyber-infrastructures for Smart Communities. The USA funding in the area of Smart Cities and Smart Communities could use a major boost in funding similar to Europe and Singapore. The New York Times article "Old World, New Tech: Europe Remains Ahead of U.S. in Creating Smart Cities" [22] points out that Europe remains ahead of USA in creating smart cities. For example, the project "The Humble Lamppost" is on the way with 30 Million Euro investment from the European Investment Bank to fund smart lampposts across EU Cities [20]. In Singapore, the National Research Foundation's Early-Stage Venture Funding Scheme announced $39 million co-funding of startups, on the private front, in 2013, venture capital invested a total of $1.71 billion in Singapore tech firms [21].

There is **a major urgency of increased funding to develop partnerships** between cities and academic and industrial partners towards establishing IoT-experimental zones and testbeds, integrations of existing IoT infrastructures and developments of new joint IoT cyber-infrastructures. The current investment towards building partnerships via the NSF Big Data Regional Hubs (BD Hubs) is a great starting point, but the funding is very small since the BD Hubs serve not only the creation of data-related partnerships for smart communities but also the creation of partnerships for other data-related societal challenges.

There is a **major urgency of continuous funding** to keep the embedded IoT cyber-infrastructures within Smart Communities up-to-date, secure and follow up with the innovations coming from IoT RD&D efforts. This is an important point and one that is quite different from many other kinds of computer science research funding. Many dimensions of the IoT solutions, which have to last decades and exist in the presence of constant technology changes, are different from traditional CS funding model. For example, the deployed IoT cyber-infrastructures for smart grid will need to last for the next 5-10 years to keep the cost of electric utility service feasible for majority of citizens. This aspect of funding is often forgotten and not planned for, causing disruptions in city services as the dependences on IoT cyber-infrastructures increase!

*For citation use*: Nahrstedt K., Lopresti D., Zorn B., Drobnis A. W., Mynatt B., Patel S., & Wright H. V. (2016). *Smart Communities Internet of Things*: A white paper prepared for the Computing Community Consortium committee of the Computing Research Association. http://cra.org/ccc/resources/ccc-led-whitepapers/

This material is based upon work supported by the National Science Foundation under Grant No. (1136993). Any opinions, findings, and conclusions or recommendations expressed in this material are those of the author(s) and do not necessarily reflect the views of the National Science Foundation.